\documentclass[prd,10pt,twocolumn,superscriptaddress,floatfix,nofootinbib]{revtex4-1}
\usepackage[utf8]{inputenc}  
\usepackage[T1]{fontenc}     
\usepackage[british]{babel}  
\usepackage[sc,osf]{mathpazo}
\usepackage[scaled=0.86]{berasans}  
\usepackage[colorlinks=true, citecolor=blue, urlcolor=blue]{hyperref} 
\usepackage{graphicx} 
\usepackage[babel]{microtype}  
\usepackage{amsmath,amssymb,amsthm,bm,amsfonts,mathrsfs,bbm} 
\DeclareMathOperator{\sech}{sech}
\usepackage{xspace}  
\usepackage{pgfplots}
\definecolor{dblue}{rgb}{0,0,0.6}
\definecolor{dred}{rgb}{1,0.08,0.58}

\begin{document}
\title{\textcolor{dblue}{Realising Mutated Hilltop Inflation in Supergravity}}
\author{Tony Pinhero}
\email{tonypinheiro2222@gmail.com}
\affiliation{Physics and Applied Mathematics Unit, Indian Statistical Institute, 203 B. T. Road, Kolkata 700108, India.}
 
\author{Supratik Pal}
\email{supratik@isical.ac.in}
\affiliation{Physics and Applied Mathematics Unit, Indian Statistical Institute, 203 B. T. Road, Kolkata 700108, India.}


\begin{abstract}
We present $\mathcal{N}=1$ supergravity models of mutated hilltop inflation (MHI) for both large and small field sectors. Models with canonical kinetic terms are developed based on a shift symmetric K\"ahler potential in inflaton superfield, and with a superpotential linear in Goldstino superfield. We also construct models with non-canonical kinetic terms for MHI by generalizing the shift symmetry. We found that a good fraction of the models can address the entire branch of MHI in a single framework.
	
\end{abstract}
\pacs{ }
\maketitle 
\section{Introduction}\label{sec_intro}

Among a plethora of models for cosmological inflation \cite{encyclopaedia_inflation_jmartin}, one of the very interesting models that  
is well-appreciated after the  latest Planck 2015 and 2018 data \cite{planck2015_inflation,planck2018_inflation}, is the
Mutated Hilltop Inflation (MHI) \cite{barun2009MHI} model. First proposed around a decade back, the salient features of these models  have been explored at length at different stages (see, e.g., \cite{barun2017MHI_revisited,barun_MHI2,encyclopaedia_inflation_jmartin}).
MHI models generically have a potential of the form 
\begin{equation}\label{MHI_potential}
V=V_{0}\left(1-\sech\mu\phi\right)
\end{equation}
for the inflaton field $\phi$ with Minkowski minimum at $\phi=0$. This model belongs to the class of chaotic inflation with super-Planckian inflaton field value $\phi\geq 1$. Depending on the value of model parameter $\mu$, MHI can occur in two branches: one belongs to large field excursion sector $\Delta\phi\geq 1$ for $\mu\lesssim 2.8$ and the other one belongs to small field sector $\Delta\phi\leq 1$ for $\mu\gtrsim2.8$ \cite{barun2017MHI_revisited}. Moreover, the model has a subset in the large field sector which is belongs to the class of $\alpha$-attractors \cite{kallosh2013universality,kallosh2013sup_alpha_attra,unity_of_cosmo_attracts,escher_in_sky,scalisi_alpha_scale,hyperbolic_geometry_of_attrctors,cosmo_attracts_nd_initial_cond_for_inflation,single_field_andre_linde,flat_alpha_attractors} for the limiting case $\mu\phi\gg1$ \cite{barun2017MHI_revisited}. Spectral index $n_{s}$ is almost independent of the parameter $\mu$, but with a slight negative running, and the tensor to scalar ratio $r$ can address the value from $10^{-4}$ to $10^{-1}$ depending upon the value of $\mu$. 
Subsequently, some more interesting features of these models have been studied in \cite{barun_MHI2,encyclopaedia_inflation_jmartin}.
These predictions are in good agreement with the  latest Planck 2015 and 2018 data \cite{planck2015_inflation,planck2018_inflation}.

However, despite all its successes, a complete description of MHI in the context of supergravity is still unavailable.
The aim of the present article is to construct a supergravity model that would lead to the mutated hilltop inflation.
 The form of the potential mentioned in Eq.(\ref{MHI_potential}) can be considered as the functional form of $\tanh \phi$ and hence it belongs to the class of $\alpha$-attractors and these $\alpha$-attractors are well formulated and studied in the context of supergravity and string theory \cite{kallosh2013universality,kallosh2013sup_alpha_attra,unity_of_cosmo_attracts,escher_in_sky,scalisi_alpha_scale,hyperbolic_geometry_of_attrctors,cosmo_attracts_nd_initial_cond_for_inflation,single_field_andre_linde,flat_alpha_attractors,seven_disc_manifold,b_mode,pole_nflation}.  In fact, the T-model
 variant of $\alpha$-attractor gives rise to a potential of the form  Eq.(\ref{MHI_potential}) and, for suitable choice  of the parameter 
  $\mu$ (and hence 
 $\alpha$), one can realize the MHI model for small-field and large-field within the framework of $\alpha$-attractor. 

The goal of this {\it letter} is to construct a supergravity framework that can account for MHI, with all non-inflaton moduli fields stabilized, and to demonstrate that it can address each and every branch of the model therefrom. 
One can accomplish this in the context of general inflaton potentials in supergravity for the chaotic inflation \cite{Kallosh2010general_inflaton_pot_in_supergravity_stabilizerfield=sGoldstino}, since MHI  falls under the category of chaotic inflation. 
In such a scenario one has to choose a K\"ahler potential which is invariant under the shift of inflaton superfield $T$ and a superpotential which is linear in Goldstino superfield $S$ \cite{kawasaki2000naturalchaotic}. More specifically, K\"ahler potential should be a function of $T\pm T^{*}$ and the component (i.e., real or imaginary part of inflaton superfield), which is not appearing in the K\"ahler potential should be treated as the real inflaton field: $T\mp T^{*}$. This is to avoid the usual $\eta$-problem in supergravity. 

Explicit functional form for these super- and K\"ahler potentials in such a construction reads
\begin{equation}\label{general_superpot_kahlerpot_in fn_form}
W=Sf(T)~~~~~~~~~K^{\pm}=K^{\pm}\left(\left(T\pm T^{*}\right)^{2},SS^{*}\right).
\end{equation}
This K\"ahler potential is invariant under the following shift transformation:
\begin{align}\label{shift_transformation}
\begin{split}
T\rightarrow T+ic~~~~~~~~\text{for}~~~ K^{+} ,
\\
T\rightarrow T+c~~~~~~~~\text{for}~~~ K^{-} .
\end{split}
\end{align}
Next step is to stabilize the Goldstino superfield $S$ at $S=0$ during inflation. This will assure the F-term SUGRA potential is positive definite via the vanishing of superpotential. i.e.,
\begin{equation}\label{poincare_sugra_potential}
V=e^{K}\left(D_{T_{i}}WK^{ij^{*}}D_{T_{j^{*}}}W^{*}-3\left|W\right|^{2}\right)\bigg|_{S=0}>0.
\end{equation}
Further, inflaton partner field $T\mp T^{*}$ will be stabilized at zero along the inflationary trajectory by attaining the mass greater than Hubble scale. As a result, the final potential will take the form
\begin{equation}
V=\left|f(T)\right|^{2}.
\end{equation} 

In the following sections we are going to elaborate on this. Specifically, we will show that our construction of canonical MHI models in supergravity are based on the above approach. Hence one can conclude that these formulations 
are the special examples of the general class of supergravity models of chaotic inflation \cite{Kallosh2010general_inflaton_pot_in_supergravity_stabilizerfield=sGoldstino}. Moreover, if one can generalize the  above shift symmetry Eq.(\ref{shift_transformation}), into the form:
\begin{equation}\label{general_shift_symmetry}
\sum_{\substack{n=1}}^{N}K^{(n)}T^{n}\rightarrow \sum_{\substack{n=1}}^{N}K^{(n)}T^{n}+C_{N}
\end{equation}
and construct a K\"ahler potential invariant under this shift, one will end up at a non-canonical kinetic term for the model. In the subsection (\ref{model-3}) we construct MHI model in supergravity in such a direction also.

\section{Mutated hilltop inflation in supergravity}
In this section, we discuss different possibilities of supergravity embedding of MHI. As mentioned above, these models are based on two superfields: one is inflaton superfield T, and another one is  Goldstino superfield S. For the standard (canonical) MHI construction we use the K\"ahler potentials $K^{\pm}$ respecting either of the shift symmetry Eq.(\ref{shift_transformation}) and representing them explicitly as follows:
\begin{equation}\label{kaehler_pot_in_canonical_form}
K^{\pm}=\pm\frac{1}{2}\left(T\pm T^{*}\right)^{2}+SS^{*}-
\zeta (SS^*)^{2}
\end{equation}
with $T\mp T^{*}$ as the inflaton. Here the term $\zeta (SS^*)^{2}$ has been added to the K\"ahler potential in order to render the mass of the stabilizer field $S$ greater than Hubble scale. So that $S$ will be stabilized during inflation. In the absence of this term, mass of the $S$ field will be light and comparable to the mass of inflaton and it will be added to the inflationary fluctuations. Hence the dynamics of the inflation cannot be regulated with single field. We explicitly discuss two branches of MHI i.e., large and small field inflation separately in the following subsections.

\subsection{Branch-I: MHI in Large Field Sector}\label{branch-1}
\subsubsection{Model-I}\label{model-1}
Let us begin with the following superpotential and K\"ahler potential 
\begin{equation}\label{superpotential1}
W=\Lambda^{2} S\frac{e^{aT/2}-e^{-aT/2}}{\left(e^{aT}+e^{-aT}\right)^{1/2}}, ~~~~~~~K=K^{-}
\end{equation}
where $a$ is the real model parameter and it should satisfy the condition $a\lesssim3.959$, in order to materialise large field inflation. Next we represent the complex superfields in terms of the real variables
\begin{equation}\label{variable_phi_chi}
T=\frac{1}{\sqrt{2}}\left(\phi+i\chi\right),~~~~~~~~~S=\frac{1}{\sqrt{2}}\left(s+i\beta\right)
\end{equation} 
where real part of $T$ is considered as inflaton. Masses of the fields along the inflationary trajectory $\chi=S=0$ is computed as follows:
\begin{multline}\label{mass_chi}
m_{\chi}^{2}=6H^{2}\left[1+\frac{a^{2}}{8}\sinh^{-2} \frac{a\phi}{2\sqrt{2}} \right.\\ \left.\times\left(1+\sech\frac{a\phi}{\sqrt{2}}-\sech^{2}\frac{a\phi}{\sqrt{2}} \right)\right]
\end{multline}
\begin{multline}\label{mass_s}
m_{S}^{2}=H^{2}\left[12\zeta+\frac{3a^{2}}{8}\sinh^{-2} \frac{a\phi}{2\sqrt{2}}\right.\\ \left.\times\sech\frac{a\phi}{\sqrt{2}}\left(1+\sech\frac{a\phi}{\sqrt{2}}\right)\right].
\end{multline}

\begin{figure}
\centering
\includegraphics[width=.9\linewidth]{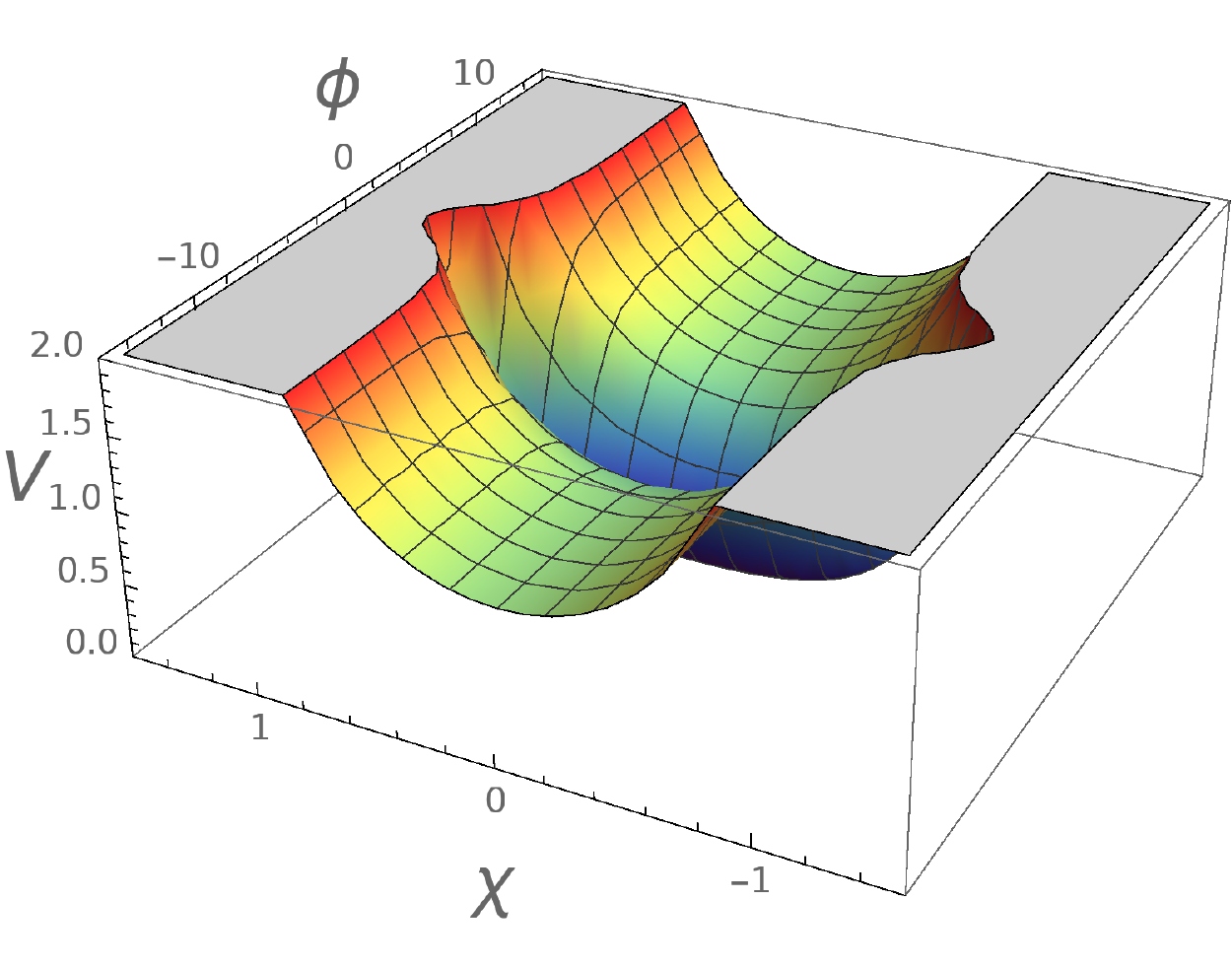}
\caption{Scalar potential for the theory Eq.(\ref{superpotential1}), for the variables Eq.(\ref{variable_phi_chi}) with $a=1$. Potential has a de Sitter valley of constant depth and width for large values of $\phi$ and has nearly Minkowski minimum at small values of $\phi$. }
\label{fig:T-model_type_pot}
\end{figure}

Thus during inflation, sinflaton attains $T-T^{*}=\chi=0$  and Goldstino $S$ stabilizes at $S=0$ for $12\zeta>1$ and the final potential, along the inflationary trajectory, reads
\begin{equation}
V|_{S=0, T-T^{*}=0}=\Lambda^{4}\left(1-\sech\frac{a\phi}{\sqrt{2}}\right)
\end{equation}
This potential is shown in the fig.(\ref{fig:T-model_type_pot}) for the field variables $\phi$ and $\chi$.

\subsubsection{Model-II}\label{model-2}
The second model in this branch can be realised by performing a change of variables $T\rightarrow iT$ in model-I,
so that the
imaginary part of the superfield $T$ is assumed as inflaton.  As a result superpotential and K\"ahler potential turn out to be
\begin{equation}
W=i\Lambda^{2} S\frac{e^{iaT/2}-e^{-iaT/2}}{\left(e^{iaT}+e^{-iaT}\right)^{1/2}}, ~~~~~~~K=K^{+}.
\end{equation}
Also, here we consider the field variables have an explicit form:
\begin{equation}\label{decompose_phi_chi}
T=\frac{1}{\sqrt{2}}\left(\chi+i\phi\right),~~~~~~~~~S=\frac{1}{\sqrt{2}}\left(s+i\beta\right).
\end{equation}

It can be readily found that, in this setup,  the inflationary dynamics and the mass matrix for the non-inflaton fields are same as that of the previous model (\ref{model-1}). With no apparent change in the above parameters but with the crucial difference of the interpretation of inflation field as above, the final potential during inflation in this model is given by
\begin{equation}
V|_{S=0, T+T^{*}=0}=\Lambda^{4}\left(1-\sech\frac{a\phi}{\sqrt{2}}\right)
\end{equation}
As apparent, this potential  will have nearly the same behavior as the one in model-I as shown in the Fig.(\ref{fig:T-model_type_pot}).

\subsubsection{Model-III}\label{model-3}
Once again we start from the model-I and perform the transformation $aT\rightarrow \sech^{-1}T$. As a result this will end up with a similar form of K\"ahler potential  and superpotential which appear in T-model supergravity $\alpha$-attractor setups \cite{hyperbolic_geometry_of_attrctors, flat_alpha_attractors}. Under the above mentioned transformation super and  K\"{a}hler potential takes the form:
\begin{equation}\label{superpotential_model3}
W=\Lambda^{2}S\sqrt{1-T},
\end{equation}
\begin{equation}\label{kaehler_potential_log_squre_1}
K=-\frac{3\alpha}{2}\log^{2}\left[\frac{T^{*}\left(1+\sqrt{1-T^{2}}\right)}{T\left(1+\sqrt{1-T^{*2}}\right)}\right]+SS^{*}-
\zeta (SS^*)^{2},
\end{equation}
where we have used $1/a=\sqrt{3\alpha}$ with $a\lesssim3.96$. Under this change of variables the new K\"ahler potential also preserves the shift symmetry. This is quite obvious from the behavior of vanishing of first term of K\"ahler potential Eq.(\ref{kaehler_potential_log_squre_1}) during inflation due to the inflaton partner $T-T^{*}$ that attains the zero vev. The K\"ahler potential Eq.(\ref{kaehler_potential_log_squre_1}) is invariant under the following shift transformation:
\begin{equation}
\log\frac{1+\sqrt{1-T^{2}}}{T}\rightarrow\log\frac{1+\sqrt{1-T^{2}}}{T}+C
\end{equation}
Corresponding kinetic term and potential for this model are, respectively, as follows:
\begin{multline}\label{kinetic_term_in _superfields}
\frac{1}{\sqrt{-g}}L_{kin}=-\frac{3\alpha}{TT^{*}\sqrt{1-T^{2}}\sqrt{1-T^{*2}}}\partial_{\mu}T
\partial^{\mu}T^{*}\\ -\left(1-4\zeta S^{*}S\right)\partial_{\mu}S\partial^{\mu}S^{*},
\end{multline}
\begin{equation}
V|_{S=0}=\Lambda^{4}e^{-\frac{3\alpha}{2}\log^{2}\left[\frac{T^{*}\left(1+\sqrt{1-T^{2}}\right)}{T\left(1+\sqrt{1-T^{*2}}\right)}\right]}\sqrt{1-T}\sqrt{1-T^{*}}
\end{equation}
Decomposing the superfields into real and imaginary parts
\begin{equation}\label{T_in_variable_phi_chi}
T=\frac{1}{\sqrt{6\alpha}}\left(\phi+i\chi\right),~~~~~~~~~S=\frac{1}{\sqrt{2}}\left(s+i\beta\right)
\end{equation}
we get the following  Lagrangian at the inflationary trajectory $S=\chi=0$:
\begin{equation}\label{lagrangian_in_real_variables}
L=\sqrt{-g}\left[\frac{R}{2}-\frac{3\alpha}{\phi^{2}\left(1-\frac{\phi^{2}}{6\alpha}\right)}\partial_{\mu}\phi\partial^{\mu}\phi-\Lambda^{4}\left(1-\frac{\phi}{\sqrt{6\alpha}}\right)\right]
\end{equation}
Further, under the field redefinition into the canonical real variable $\psi$: $\phi=\sqrt{6\alpha}\sech(\psi/\sqrt{6\alpha})$, the final Lagrangian reads:
\begin{equation}
L=\sqrt{-g}\left[\frac{R}{2}-\frac{1}{2}\partial_{\mu}\psi\partial^{\mu}\psi-\Lambda^{4}\left(1-\sech\frac{\psi}{\sqrt{6\alpha}}\right)\right].
\end{equation}
\begin{figure}
	\centering
	\includegraphics[width=.9\linewidth]{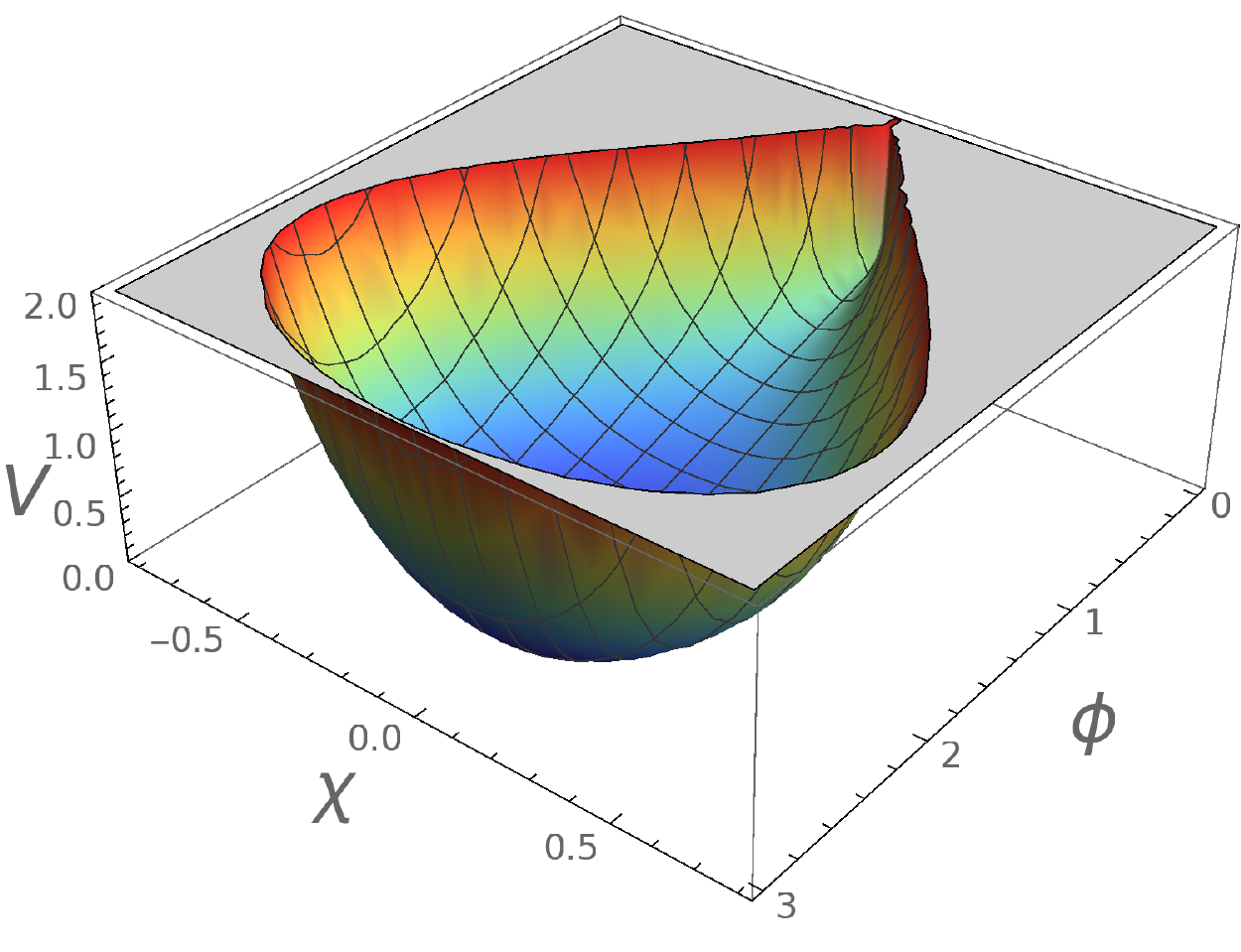}
	\caption{Scalar potential for the theory Eq.(\ref{superpotential_model3}) and  Eq.({\ref{kaehler_potential_log_squre_1}}), for the variables Eq.(\ref{T_in_variable_phi_chi}). Although the existence of flat direction in the potential is not obvious in this set of variables, there indeed is a flat direction as demonstrated in Sec.\ref{model-3}.}
	\label{fig:pot_in_non-canonical_variables}
\end{figure}

The scalar potential for the model Eq.(\ref{kaehler_potential_log_squre_1}) and Eq.(\ref{superpotential_model3}) is shown in Fig.(\ref{fig:pot_in_non-canonical_variables}). One may wonder if the flat direction is visible in these non-canonical variables $\phi$ and $\chi$. 
In order to clarify that, we represent them in a more convenient variable $\hat{T}=(\psi+i\theta)\sqrt{6\alpha}$, which is related to $T$ as $T=\sech\hat{T}$. Consequently,  the kinetic terms become canonical for both fields $\psi$ and $\theta$ and the potential will look approximately same as that of model-I as shown in the Fig.(\ref{fig:T-model_type_pot}), thereby reassuring  flat directions.

Nevertheless, the moduli space associated with this model, i.e., K\"ahler manifold associated with the K\"ahler potential Eq.(\ref{kaehler_potential_log_squre_1}) is geometrically flat. This can be demonstrated as follows. The metric of the moduli space based on the K\"ahler potential Eq.(\ref{kaehler_potential_log_squre_1}) is defined as

\begin{equation}
ds^{2}=g_{TT^{*}}dT dT^{*}
\end{equation}
where
\begin{equation}
g_{TT^{*}}=K_{TT^{*}}=-\frac{3\alpha}{TT^{*}\sqrt{1-T^{2}}\sqrt{1-T^{*2}}}.
\end{equation}
From this K\"ahler metric non-vanishing Levi-Civita connection coefficients, Riemannian tensors, and the curvature of the moduli space are computed as follows:
\begin{equation}
\Gamma_{TT}^{T}=-\frac{1-2T^{2}}{T-T^{3}},~~~~~~\Gamma_{T^{*}T^{*}}^{T^{*}}=-\frac{1-2T^{*2}}{T^{*}-T^{*3}}
\end{equation}
\begin{equation}
\mathcal{R}_{TT^{*}T}^{T}=\partial_{T^{*}}\Gamma_{TT}^{T}=0.
\end{equation}
\begin{equation}
\mathcal{R}_{\text{K\"ahler}}=0
\end{equation}
 Alternatively, from the definition of curvature of K\"ahler manifold via the metric one can straightaway show that
\begin{equation}
\mathcal{R}_{\text{K\"ahler}}=-g_{\Phi\Phi^{*}}^{-1}\partial_{\Phi}\partial_{\Phi^{*}}\log g_{\Phi\Phi^{*}}=0.
\end{equation}
From the above, one can conclude that geometry associated with our K\"ahler manifold is indeed flat.

\subsubsection{Model-IV}
Till now, we were discussing about MHI models in supergravity based on a single parameter. A two parameter realisation of the scenario in supergravity is also possible for MHI model. In such a scenario one can consider a combination of superpotential and K\"ahler potential as follows:
\begin{equation}\label{superpotential_two_parameter}
W=\Lambda^{2} S\frac{e^{-aT/2}-e^{-bT/2}}{\left(e^{-aT}+e^{-bT}\right)^{1/2}}, ~~~~~~~K=K^{-}
\end{equation}
This will end up with a potential of the form for the real part of the inflaton superfield $T$:
\begin{equation}\label{potential_two_parameter}
V|_{S=0, T-T^{*}=0}=\Lambda^{4}\left(1-\sech\frac{(a-b)\phi}{2\sqrt{2}}\right)
\end{equation}
for the variables defined in Eq.(\ref{variable_phi_chi}). Thus, although we started with two parameters in supergravity, the form of the potential boils down to  such a form so as to be represented by a single parameter, namely, $(a-b)$. As a result, in this model the parameters $a$ and $b$ can take any value so far as the constraint $-7.191\lesssim (a-b) \lesssim 7.191$ is satisfied in order to guarantee large field excursion. The mass matrix for the non-inflaton fields along the inflationary trajectory are computed as follows:
\begin{multline}\label{mass_chi_two_variable_model}
m_{\chi}^{2}=6H^{2}\left[1+\frac{(a-b)^{2}}{16}\left(\sinh^{-2} \frac{(a-b)\phi}{4\sqrt{2}}\right.\right.\\ \left.\left.+2\sech^{2}\frac{(a-b)\phi}{2\sqrt{2}}\right) \right]
\end{multline}
\begin{multline}\label{mass_s_two_variable_model}
m_{S}^{2}=H^{2}\left[12\zeta+\frac{3(a-b)^{2}}{32}\sinh^{-2} \frac{(a-b)\phi}{4\sqrt{2}}\right.\\ \left.\times\sech\frac{(a-b)\phi}{2\sqrt{2}}\left(1+\sech\frac{(a-b)\phi}{2\sqrt{2}}\right)\right].
\end{multline}
The same will also work   for the field replacement of all scalars of the type $Z\rightarrow iZ$ in Eq.(\ref{superpotential_two_parameter}) (as in model-(\ref{model-2})). In that case the super- and K\"ahler potentials take the form:
\begin{equation}\label{superpotential_two_parameter_in_i}
W=i\Lambda^{2} S\frac{e^{-iaT/2}-e^{-ibT/2}}{\left(e^{-iaT}+e^{-ibT}\right)^{1/2}}, ~~~~~~~K=K^{+}
\end{equation}
where the inflaton field will be the imaginary part of the superfield $T$ and it will end up with the potential Eq.(\ref{potential_two_parameter}) for the variables Eq.(\ref{decompose_phi_chi}).

\subsection{Branch-II: MHI in Small Field Sector}\label{small_field_model}
\subsubsection{Model-I}
\begin{figure}
	\centering
	\includegraphics[width=.9\linewidth]{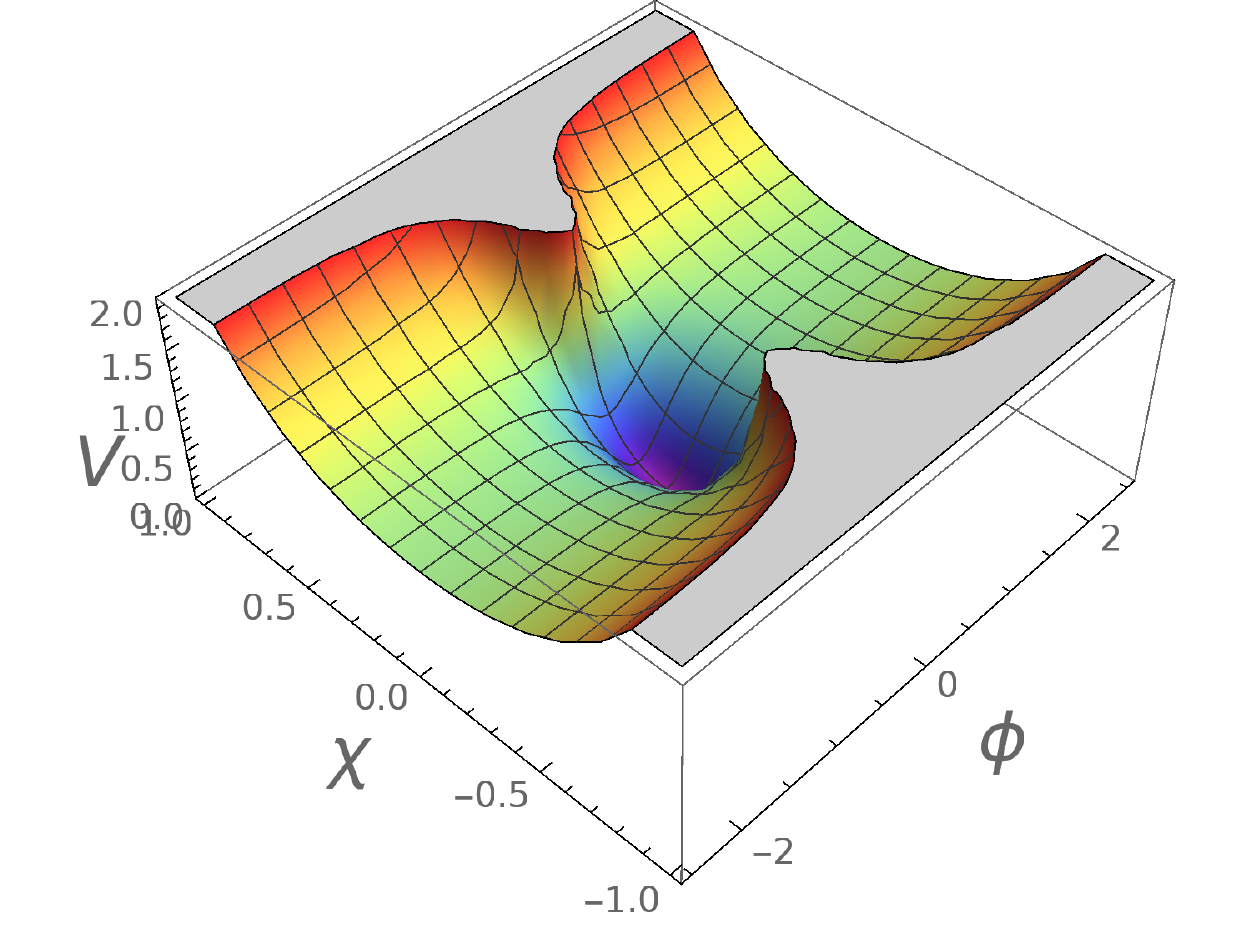}
	\caption{Scalar potential for the theory Eq.(\ref{superpotential1}), for the variables Eq.(\ref{variable_phi_chi}) with $a=4.242$. This potential also has a de Sitter valley of constant depth and width for large values of $\phi$ and has nearly Minkowski minimum at small values of $\phi$.}
	\label{fig_small_field_potential}
\end{figure}

The first kind of models in Small Field Sector is rather trivial.
It is a straightforward exercise to show that
all the models  presented in  section-(\ref{branch-1}) in the context of large field inflation can also govern small field excursion if  the value of the model parameter(s) $a\gtrsim3.959~~\text{or}~~(a-b)\gtrsim 7.191$. In this scenario all non-inflaton moduli fields are stabilized in the inflationary trajectory with a slightly different mass. This is quite evident from the expressions of mass matrix of the fields Eq.(\ref{mass_chi}) and Eq.(\ref{mass_s}). There is a small change in the shape of the potential along $\chi$ direction at small $\phi$ region.
We have demonstrated the results  in  Fig.(\ref{fig_small_field_potential}) that reflects distinctive features of small field models.

\subsubsection{Model-II}
Until now, we have supersymmetric  models which can simultaneously explain both large field and small field MHI. In what follows we will construct an exclusive model for small field MHI.  For this let us consider super- and K\"ahler potentials as:
\begin{equation}\label{superpotential_small_field_non-canonical}
W=\Lambda^{2}Se^{\frac{3}{2}\alpha\log^{2}\left(\frac{T}{1+\sqrt{1-T^{2}}}\right)}\sqrt{1+T},
\end{equation}
\begin{equation}\label{kahler_pot_small_field_non-canonical}
K=-3\alpha\left | \log\left(\frac{T}{1+\sqrt{1-T^{2}}}\right) \right |^{2}+SS^{*}-
\zeta (SS^*)^{2}.
\end{equation}
Where $\alpha\lesssim 0.0212$ to satisfy the condition of small field inflation. The associated kinetic and potential terms for this model are given by
\begin{multline}\label{kinetic_term_in _superfields_small_field}
\frac{1}{\sqrt{-g}}L_{kin}=\frac{3\alpha}{TT^{*}\sqrt{1-T^{2}}\sqrt{1-T^{*2}}}\partial_{\mu}T
\partial^{\mu}T^{*}\\ -\left(1-4\zeta S^{*}S\right)\partial_{\mu}S\partial^{\mu}S^{*},
\end{multline}
\begin{multline}
V=\Lambda^4 e^{\frac{3}{2}\alpha\left[\log^{2}\left(\frac{T}{1+\sqrt{1-T^{2}}}\right)+\log^{2}\left(\frac{T^{*}}{1+\sqrt{1-T^{*2}}}\right)\right]}\\e^{-3\alpha\left | \log\left(\frac{T}{1+\sqrt{1-T^{2}}}\right) \right |^{2}} \times \sqrt{1-T}\sqrt{1-T^{*}}
\end{multline}
where the potential is calculated at $S=0$. Decomposing these superfields into real and imaginary parts
\begin{equation}\label{T_in_variable_phi_chi_small_field}
T=\frac{1}{\sqrt{6\alpha}}\left(\phi+i\chi\right),~~~~~~~~~S=\frac{1}{\sqrt{2}}\left(s+i\beta\right)
\end{equation}
the total Lagrangian at the inflationary trajectory $S=\chi=0$ takes the form
\begin{equation}\label{lagrangian_in_real_variables_small_field}
L=\sqrt{-g}\left[\frac{R}{2}+\frac{3\alpha}{\phi^{2}\left(1-\frac{\phi^{2}}{6\alpha}\right)}\partial_{\mu}\phi\partial^{\mu}\phi-\Lambda^{4}\left(1-\frac{\phi}{\sqrt{6\alpha}}\right)\right]
\end{equation}
Consequently, under the following field redefinition in the canonical real variable $\psi$: $\phi=\sqrt{6\alpha}\sech(\psi/\sqrt{6\alpha})$, the final Lagrangian reads:
\begin{equation}
L=\sqrt{-g}\left[\frac{R}{2}-\frac{1}{2}\partial_{\mu}\psi\partial^{\mu}\psi-\Lambda^{4}\left(1-\sech\frac{\psi}{\sqrt{6\alpha}}\right)\right].
\end{equation}
Which represents the mutated hilltop inflation (MHI).
Scalar potential for this theory is shown in the fig (\ref{fig:potential_small_field_non_canonical}) in terms of the more adequate variables $\hat{T}=(\psi+i\theta)\sqrt{6\alpha}$, which is related to $T$ as $T=\sech\hat{T}$. The canonical masses for all those non-inflaton stabilized fields are computed in terms of the variable $\hat{T}$ along the inflationary trajectory as follows:
\begin{figure}
	\centering
	\includegraphics[width=.9\linewidth]{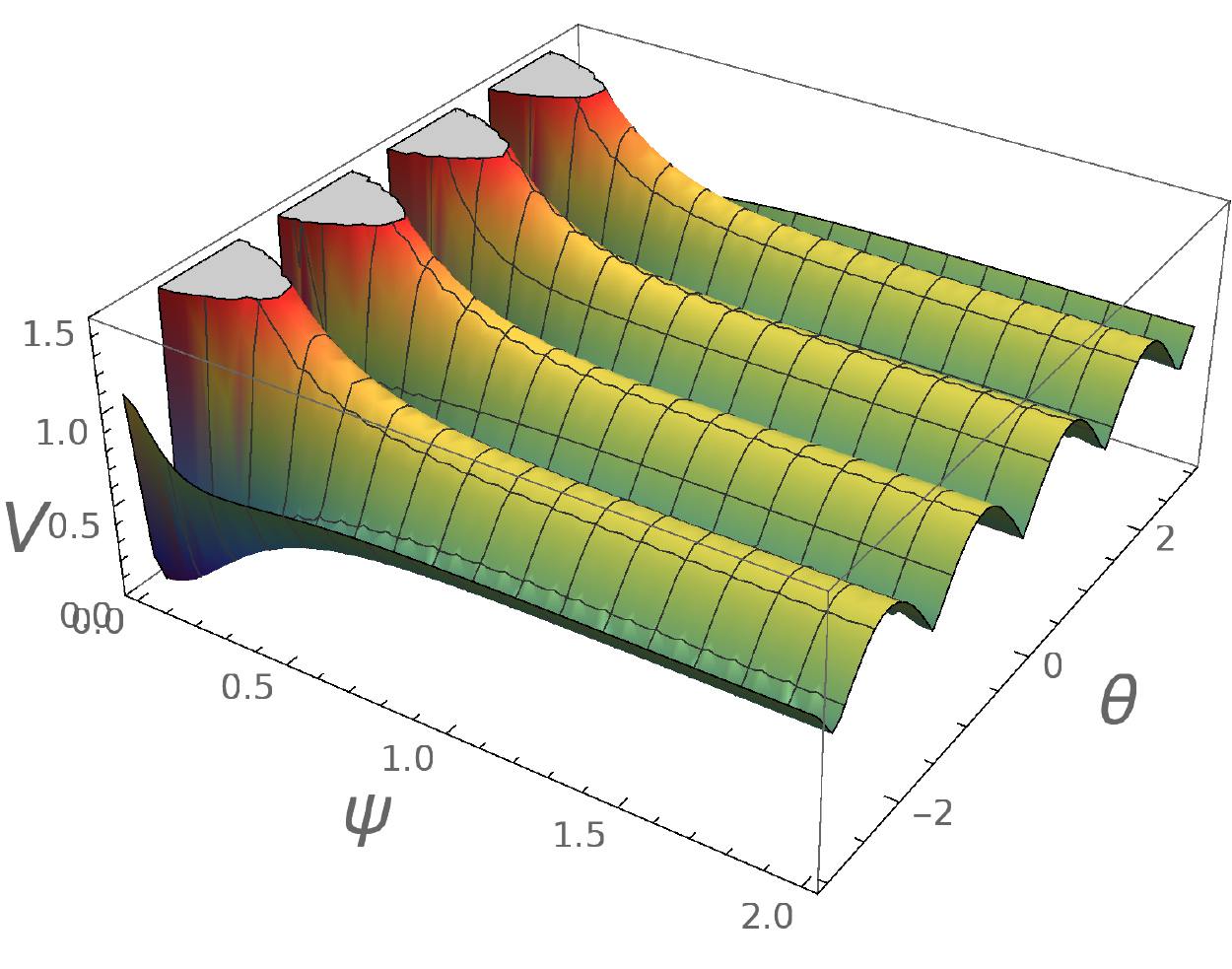}
	\caption{Scalar potential for the theory Eq.(\ref{superpotential_small_field_non-canonical}) and  Eq.(\ref{kahler_pot_small_field_non-canonical}) for $\alpha=0.008$.}
	\label{fig:potential_small_field_non_canonical}
\end{figure}
\begin{multline}\label{mass_chi_sfmhi}
m_{\theta}^{2}=6H^{2}\left[1-\frac{1}{24\alpha}\left(\sinh \frac{\psi}{2\sqrt{6\alpha}}\right)^{-2} \right.\\ \left.\times\left(1+\sech\frac{\psi}{\sqrt{6\alpha}}-\sech^{2}\frac{\psi}{\sqrt{6\alpha}} \right)\right]
\end{multline}
\begin{multline}\label{mass_s_sfmhi}
m_{S}^{2}=H^{2}\left[12\zeta-\frac{1}{4\alpha}\sinh^{-2} \frac{\psi}{\sqrt{6\alpha}}\right.\\ \left.\times\sech\frac{\psi}{\sqrt{6\alpha}}\left(1+\sech\frac{\psi}{\sqrt{6\alpha}}\right)\right].
\end{multline}

\section{Concluding remarks}
In this {\it letter}, we have presented various supergravity embedding of mutated hilltop inflation (MHI) model for both large and small field branches with stable Minkowski vacuum. Models constructed with canonical kinetic terms are the special examples of general class of models describing chaotic inflation in supergravity \cite{Kallosh2010general_inflaton_pot_in_supergravity_stabilizerfield=sGoldstino}. These canonical models are developed based on a shift symmetric K\"ahler potential in inflaton superfield, and with a superpotential linear in Goldstino superfield. By generalizing this shift symmetry, we also constructed models for MHI with non-canonical kinetic terms. We also found that many of the  models described in the {\it letter} can address the entire branch (large field and small field sectors) of MHI in a single framework.


\section*{Acknowledgments}

TP would like to  thank Barun Kumar Pal and Abhishek Naskar for the enlightening discussions. T.P gratefully acknowledge the support from Senior Research fellowship (Order No. DS/18-19/0616) of the Indian Statistical Institute (ISI), Kolkata.




\begin{thebibliography}{99}
\bibitem{encyclopaedia_inflation_jmartin}J. Martin, C. Ringeval, V. Vennin, \emph{Encyclopaedia Inflationaris}, \href{https://www.sciencedirect.com/science/article/pii/S2212686414000053?via%3Dihub}{ Phys.Dark Univ. 5-6 (2014) 75-235}, \href{https://arxiv.org/abs/1303.3787}{\color{dred}[arXiv:1303.3787 [astro-ph.CO]]}	



\bibitem{planck2015_inflation}Planck Collaboration, P. A. R. Ade et al., \emph{Planck 2015 results. XX. Constraints on inflation}, 	\href{https://www.aanda.org/articles/aa/abs/2016/10/aa25898-15/aa25898-15.html}{Astron.Astrophys. 594 (2016) A20},  \href{https://arxiv.org/abs/1502.02114}{\color{dred}[arXiv:1502.02114 [astro-ph.CO]]}



\bibitem{planck2018_inflation}Planck Collaboration, Y. Akrami et al., \emph{Planck 2018 results. X. Constraints on inflation}, \href{https://arxiv.org/abs/1807.06211}{\color{dred}[ arXiv:1807.06211 [astro-ph.CO]]} 


	
	
	
\bibitem{barun2009MHI}B. K. Pal, S. Pal, B. Basu, \emph{Mutated Hilltop Inflation : A Natural Choice for Early Universe}, \href{https://iopscience.iop.org/article/10.1088/1475-7516/2010/01/029/meta}{JCAP 1001 (2010) 029}, \href{https://arxiv.org/abs/0908.2302}{\color{dred}[arXiv:0908.2302 [hep-th]]}	
	

\bibitem{barun2017MHI_revisited}B. K. Pal, \emph{Mutated Hilltop Inflation Revisited}, \href{https://link.springer.com/article/10.1140%2Fepjc%2Fs10052-018-5856-3}{Eur.Phys.J. C78 (2018) no.5, 358}, \href{https://arxiv.org/abs/1711.00833}{\color{dred}[arXiv:1711.00833 [gr-qc]]}

\bibitem{barun_MHI2} B. K. Pal, S. Pal, B. Basu,\emph{A semi-analytical approach to perturbations in mutated hilltop inflation}, \href{https://www.worldscientific.com/doi/abs/10.1142/S0218271812500174}{Int.J.Mod.Phys. D21 (2012) 1250017}, \href{https://arxiv.org/abs/1010.5924}{\color{dred}[arXiv:1010.5924 [astro-ph.CO]]}



\bibitem{kallosh2013universality} R. Kallosh and A. Linde, \emph{Universality Class in Conformal Inflation}, 
\href{http://iopscience.iop.org/article/10.1088/1475-7516/2013/07/002/meta;jsessionid=0C76449ABC7D9A14139309271C43F61E.c2.iopscience.cld.iop.org}{JCAP 1307 (2013) 002},
\href{https://arxiv.org/abs/1306.5220}{\color{dred}[arXiv:1306.5220 [hep-th]]}



\bibitem{kallosh2013sup_alpha_attra}R. Kallosh, A. Linde, D. Roest, \emph{Superconformal Inflationary $\alpha$-Attractors}, \href{http://link.springer.com/article/10.1007%2FJHEP11%282013%29198}{JHEP 1311 (2013) 198}, \href{https://arxiv.org/abs/1311.0472}{\color{dred}[arXiv:1311.0472 [hep-th]]}	
	
\bibitem{unity_of_cosmo_attracts}M. Galante, R. Kallosh, A. Linde, D. Roest, \emph{The Unity of Cosmological Attractors}, \href{https://journals.aps.org/prl/abstract/10.1103/PhysRevLett.114.141302}{Phys.Rev.Lett. 114 (2015) no.14, 141302}, \href{https://arxiv.org/abs/1412.3797}{\color{dred}[arXiv:1412.3797 [hep-th]]}

\bibitem{escher_in_sky}R. Kallosh and A. Linde, \emph{Escher in the Sky}, \href{https://www.sciencedirect.com/science/article/pii/S1631070515001309?via%3Dihub}{Comptes Rendus Physique 16 (2015) 914-927}, \href{https://arxiv.org/abs/1503.06785}{\color{dred}[arXiv:1503.06785 [hep-th]]}



\bibitem{scalisi_alpha_scale}
D.Roest, M. Scalisi,  \emph{Cosmological Attractors from $\alpha$-Scale Supergravity}, \href{https://journals.aps.org/prd/abstract/10.1103/PhysRevD.92.043525}{Phys.Rev. D92 (2015) 043525}, \href{https://arxiv.org/abs/1503.07909}{\color{dred}[arXiv:1503.07909 [hep-th]]}		



\bibitem{hyperbolic_geometry_of_attrctors}J. J. M. Carrasco, R. Kallosh, A. Linde, D. Roest, \emph{The Hyperbolic Geometry of Cosmological Attractors}, \href{https://journals.aps.org/prd/abstract/10.1103/PhysRevD.92.041301}{ Phys.Rev. D92 (2015) no.4, 041301}, \href{https://arxiv.org/abs/1504.05557}{\color{dred}[arXiv:1504.05557 [hep-th]]}

\bibitem{cosmo_attracts_nd_initial_cond_for_inflation}J. J. M. Carrasco, R. Kallosh, A. Linde, \emph{Cosmological Attractors and Initial Conditions for Inflation}, \href{https://journals.aps.org/prd/abstract/10.1103/PhysRevD.92.063519}{Phys.Rev. D92 (2015) no.6, 063519}, \href{https://arxiv.org/abs/1506.00936}{\color{dred}[arXiv:1506.00936 [hep-th]]}

\bibitem{single_field_andre_linde} A. Linde \emph{Single-field $\alpha$-attractors}, \href{http://iopscience.iop.org/article/10.1088/1475-7516/2015/05/003/meta}{JCAP 1505 (2015) 003} \href{https://arxiv.org/abs/1504.00663}{\color{dred}[arXiv:1504.00663 [hep-th]]}

\bibitem{flat_alpha_attractors} T. Pinhero, \emph{Natural $\alpha$-Attractors from ${\cal N}=1$ Supergravity via flat K\"ahler Manifolds}, \href{https://arxiv.org/abs/1812.05406}{\color{dred}[arXiv:1812.05406 [hep-th]]}








	
\bibitem{seven_disc_manifold}Sergio Ferrara and Renata Kallosh, \emph{Seven-Disk Manifold, alpha-attractors and B-modes}, \href{https://journals.aps.org/prd/abstract/10.1103/PhysRevD.94.126015}{Phys.Rev. D94 (2016) no.12, 126015}, \href{https://arxiv.org/abs/1610.04163}{\color{dred}[arXiv:1610.04163 [hep-th]]}




\bibitem{b_mode}R. Kallosh, A. Linde, T. Wrase, Y. Yamada, \emph{Maximal Supersymmetry and B-Mode Targets}, \href{https://link.springer.com/article/10.1007%2FJHEP04%282017%29144}{JHEP 1704 (2017) 144}, \href{https://arxiv.org/abs/1704.04829}{\color{dred}[arXiv:1704.04829 [hep-th]]}
	
	
\bibitem{pole_nflation}M. Dias, J. Frazer, A. Retolaza, M. Scalisi, A. Westphal, \emph{Pole N-flation}, \href{https://arxiv.org/abs/1805.02659}{\color{dred}[arXiv:1805.02659 [hep-th]]}
	
	
	
	

\bibitem{Kallosh2010general_inflaton_pot_in_supergravity_stabilizerfield=sGoldstino}R. Kallosh, A. Linde, T. Rube, \emph{General inflaton potentials in supergravity}, \href{http://journals.aps.org/prd/abstract/10.1103/PhysRevD.83.043507}{Phys.Rev.D83:043507,2011}, \href{https://arxiv.org/abs/1011.5945}{\color{dred}[arXiv:1011.5945 [hep-th]]}



\bibitem{kawasaki2000naturalchaotic}M. Kawasaki, M. Yamaguchi, T. Yanagida, \emph{Natural Chaotic Inflation in Supergravity}, \href{http://journals.aps.org/prl/abstract/10.1103/PhysRevLett.85.3572}{Phys.Rev.Lett. 85 (2000) 3572-3575},  \href{https://arxiv.org/abs/hep-ph/0004243}{\color{dred}[arXiv:hep-ph/0004243]}





	

	

		
	


		
		

			
			











\end{thebibliography}
\end{document}